\documentclass[preprint,review,12pt,1p,times]{elsarticle}

\usepackage{amsmath,amssymb,amsfonts,quotes}
\usepackage{algorithmic}
\usepackage{graphicx}
\usepackage[ruled,vlined,linesnumbered]{algorithm2e}
\usepackage{hyperref}
\usepackage{textcomp}
\usepackage{tabularx,booktabs}
\usepackage{tikz, pgfplots}
\usepackage{caption}
\usepackage{subcaption}
\usepackage{float}
\restylefloat{table}
\pgfplotsset{compat=1.18}
\usetikzlibrary{positioning}
\graphicspath{ {pictures/} }

\def\BibTeX{{\rm B\kern-.05em{\sc i\kern-.025em b}\kern-.08em
		T\kern-.1667em\lower.7ex\hbox{E}\kern-.125emX}}

\journal{Applied Soft Computing}

\newcolumntype{Y}{>{\centering\arraybackslash}X}

\newcommand{\drawHACgeneralStructure}{
	\begin{figure}
		\centering
		\normalsize
		\begin{tikzpicture}[
			BLOCK_STYLE/.style={rectangle, draw=black!60, fill=black!1, very thick, minimum size=5mm},
			CONTAINER_STYLE/.style={rectangle, draw=blue!60, fill=none, very thick, minimum width=4cm, minimum height=4cm, dashed},
			SYSTEM_STYLE/.style={rectangle, draw=red!100, fill=red!5, very thick, minimum size=10mm},
			]
			\node[CONTAINER_STYLE](Container)[]{};
			\node[BLOCK_STYLE](Semantization)[above=1.3cm of Container.center]{Semantization};
			\node[BLOCK_STYLE](HA Inference)[below=of Semantization]{HA Inference};
			\node[BLOCK_STYLE](De-Semantization)[below=of HA Inference]{De-Semantization};
			\node[BLOCK_STYLE](Adaptive Weights)[below=of De-Semantization]{Weighted combination};
			\node[SYSTEM_STYLE](System)[right=12mm of HA Inference]{System};
			\node(Notation)[above=2mm of Semantization]{\(i=1,2,...,n\)};
			
			\draw[->, very thick, dashed] (Semantization.south) to node[right] {\(x_{is}\)} (HA Inference.north);
			\draw[->, very thick, dashed] (HA Inference.south) to node[right] {\(u_{is}\)} (De-Semantization.north);
			\draw[->, very thick] (De-Semantization.south) to node[right] {\(u_{i}\)} (Adaptive Weights.north);
			\draw[->, very thick] (Adaptive Weights.east) -| node[right] {\(u_j,\,j=1,2,...,m\)}(System.south);
			\draw[->, very thick] (System.north) |- node[right] {\(x_i\)} (Semantization.east);
		\end{tikzpicture}
		\caption{HAC general structure for a system with \(n\) observable states and \(m\) inputs. Each state is separately inferred to get its corresponding intermediate control action. Subsequently, all theses actions are weighted, then combined into \(m\) final control signals. Dashed arrows represents variables in semantic domains, otherwise they are in real domains.}\label{HAC_struct}
	\end{figure}
}

\begin{document}
	
	\begin{frontmatter}
		
		
		
		\title{Design and Implementation of Hedge Algebra Controller using Recursive Semantic Values for Cart-pole System}
		
		
		\author{Nhat-Minh Dzoan$^a$} 
		\ead{minh.dn195793@sis.hust.edu.vn}
		\affiliation{organization={School of Mechanical Engineering, Hanoi University of Science and Technology},
			addressline={No. 1 Dai Co Viet, Back Khoa, Hai Ba Trung}, 
			city={Hanoi},
			country={Vietnam}}
			
		\author{Thi-Thoa Mac*,$^a$} 
		\ead{Coresponding author: thoa.macthi@hust.edu.vn}
		\author{Hoang-Hiep Ly$^a$} 
		\ead{hiep.lyhoang@hust.edu.vn}
		\author{Xuan-Thuan Nguyen$^a$} 
		\ead{thuan.nguyenxuan@hust.edu.vn}
		\begin{abstract}
			This paper presents a novel approach to designing a Hedge Algebra Controller named Hedge Algebra Controller with Recursive Semantic Values (RS-HAC). This approach incorporates several newly introduced concepts, including Semantically Quantifying Simplified Mapping (SQSM) featuring a recursive algorithm, Infinite General Semantization (IGS), and Infinite General De-semantization (IGDS). These innovations aim to enhance the optimizability, scalability, and flexibility of hedge algebra theory, allowing the design of a hedge algebra-based controller to be carried out more efficiently and straightforward. An application of stabilizing an inverted pendulum on a cart is conducted to illustrate the superiority of the proposed approach. Comparisons are made between RS-HAC and a fuzzy controller of Takagi-Sugeno type (FC), as well as a linear quadratic regulator (LQR). The results indicate that the RS-HAC surpasses the FC by up to 400\%  in control efficiency and is marginally better than the LQR regarding transient time in balancing an inverted pendulum on a cart.
		\end{abstract}
		
		
	\end{frontmatter}
		
	\section{Introduction}
	\label{intro}
Fuzzy set theory had its debut in 1965\cite{ZADEH1965338}, and has long been applied to design controllers for mechanical systems as it offers a way of modeling human reasoning process, knowledge and experience mathematically. Therefore, fuzzy controllers can be designed solely based on the expert's knowledge and experience without deriving an explicit dynamical model of the plant. In\cite{LIU2024111818}, provides a comprehensive review of the integration of fuzzy theory with natural language processing (NLP). It discusses the current limitations of this fusion and addresses future challenges while offering suggestions for potential research topics.  Hedge algebra (HA) theory was first introduced in 1999 as an extension to fuzzy set theory, delivering a reasoning method that directly handles linguistic terms and an algebraic metric that allows maintaining the semantic order of linguistic variables \cite{doi:10.1142/S0218488599000301,HO2008968}. Since then, HA-based controllers have gradually gained popularity because the fuzzy inference process can be simplified as an interpolation on pre-computed surfaces or lines. This removes the need for a de-fuzzification process and all the computational-costly membership functions. Thus, extremely low CPU time can be achieved, while comparable performance is also attained \cite{sqm_interpolation_2006,intermediate_ctrl_2022}.
\par
The application of hedge algebra in control theory has been extensively studied since 2007. In \cite{HO2008968}, a comprehensive workflow of designing a hedge algebra controller (HAC) is presented with an extra step of applying a genetic algorithm to find optimal parameters for the controller. This controller is then applied to solve the inverted pendulum problem and the problem of holding an object on a "hill". In \cite{HAC_earthquake_2013}, a practical application of HAC is presented. The controller is designed to actively dampen a 10-story building's vibration under the influence of an earthquake. A hedge algebra inference engine based on the semantic associative memory (SAM) table is used to represent the reasoning process that deduces the control input from the current states of the building. In \cite{intermediate_ctrl_2022}, another HAC structure is demonstrated to solve the problem of balancing a cart pole. This HAC separately infers multiple intermediate controllers for each system's state and calculates the overall control action as an adaptive weighted sum of all intermediate actions. This type of HAC, inspired by the Takagi-Sugeno fuzzy model\cite{6313399} and the single input rule modules (SIRMs) \cite{YI2000153}, is a big improvement compared to previous models since it makes the design process more straightforward by decoupling system inputs in each rule implication, thus reducing the number of control rules significantly. In \cite{doi:10.1061/JENMDT.EMENG-6821}, a novel device called "active upgraded tuned liquid column damper" (AUTLCD) controlled by a HA-based controller, is introduced. This device is installed on a high-rise building to minimize its structural vibration when the structure is under the influence of an earthquake. The AUTLCD with its HAC is optimized by a recently developed optimization scheme called "balancing composite motion optimization" (BCMO) \cite{BCMO}. The proposed device is potentially applicable to real-world scenarios as it shows promising results when applied to the well-known El Centro earthquake data in 1940. Recently, hedge algebra-based controllers have been developed for robotics applications. An example is in \cite{traj_track_HAC}, where the BCMO is once again utilized to find the optimal parameters for a HAC (oHA controller) that controls a generic differential drive robot tracking a reference trajectory. The oHA controller outperforms a fuzzy-based controller regarding trajectory tracking ability, calculation time, and robustness in various reference trajectories.
\par
Although hedge algebra theory provides a solid foundation for designing effective and model-free controllers suitable for many applications, it could be better. One noticeable disadvantage of HA theory is the process of calculating "semantically quantifying mapping" (SQM) values for all the linguistic labels \cite{HO2008968}. Remarkably, the formulation of SQM (shown later in section \ref{HA_theory}) consists of a special \textit{Sign} function \cite{HO2008968, sqm_interpolation_2006} that requires human's assessment to compute. This results in experts who know the system at hand being involved in calculating the SQM values for the plant. Otherwise, these values are determined through trials and errors, or even for popular systems; these values are pre-computed in a look-up table by other forerunners such as in \cite{traj_track_HAC}.
Moreover, the lengthy and complicated formulation of SQM \cite{HO2008968} leads to the fact that every SQM of each linguistic label has different and non-recursive formulas, inherently creating a scalability issue. Thereby, it is cumbersome to change the number of linguistic labels or hedge elements in the middle of the operation without reformulating the whole array of SQM formulas, making the optimization process of the controller very limited and challenging. Lastly, the HA sanitization or de-sanitization process requires prior knowledge about the crisp domain of a state variable (i.e., the real numerical boundary of a state of the control plant). However, for the wide variety of real-world systems, it is only possible to acquire this information after running the controller (not to mention designing it). Let us take an under-actuated plant as a case in point. The boundaries of some states of this type of plant cannot be pre-determined. Therefore, trial and error is the current technique to deal with these systems (i.e., considering the state boundaries as optimization variables).
\par
In this study, we aim to introduce a new genre of hedge algebra-based controllers called "hedge algebra controller with recursive semantic values" (RS-HAC) that incorporates two freshly proposed concepts: "semantically quantifying simplified mapping" (SQSM) and "infinite general sanitization" (IGS). In particular, SQSM is the simplified and upgraded version of SQM. It features a recursive and programmable formulation that eliminates the need for human assessment of the semantic degree of hedge elements, thereby addressing all fundamental problems inherent in SQM. Moreover, SQSM is easily scalable and flexible because it allows all formulation parameters to change dynamically, removing all SQM limitations regarding how the controller can be optimized. On the other hand, the IGS proposed in this paper makes the sanitization and de-sanitization of the hedge algebra theory more intuitive and versatile as it is a general mapping that does not require prior knowledge about the crisp domain of a system state. Together, these two concepts are crucial additions to the foundation of hedge algebra theory, contributing to its further refinement.
\par
Hereafter, we outline some main features of the traditional HA theory in section \ref{HA_summary}. The proposed RS-HAC, together with the SQSM and the IGS, are presented in section \ref{RSHAC}, followed by their application in solving the cart-pole problem and comparison with a fuzzy controller and a linear quadratic regulator in section \ref{app_exp}.
\section{Summary of Hedge algebra theory}\label{HA_summary}
	Below is the brief summary of the theoretical basis of hedge algebra based on \cite{doi:10.1142/S0218488599000301, HO2008968, sqm_interpolation_2006, intermediate_ctrl_2022}. Readers may refer to \cite{HO2008968, sqm_interpolation_2006} for detailed explanations.
	\subsection{The hedge algebra theory}\label{HA_theory}
	The differences separating hedge algebra from fuzzy set theory is that an HA structure maintains its own semantic based arrangement among all linguistic labels of a given linguistic variable. For example, consider the linguistic variable \textbf{SIZE}, let us derive 7 vague linguistic labels from this variable which are \textbf{very small, small, little small, neutral, little big, big, very big}. As a matter of common sense, we can easily arranged these terms based on their semantic meanings as follows: \textbf{$\text{very small} < \text{small} < \text{little small} < \text{neutral} < \text{little big} < \text{big} < \text{very big}$}. Hedge algebra is a mathematical tool that can maintain these relations algebraically by quantifying their semantic values using a mapping called "Semantically quantifying mapping" (SQM). Let us show this example in hedge algebra fashion.
	
	The HA structure of the linguistic variable \textbf{SIZE} is expressed as follows:
	\begin{align}
		AX=(X,G,C,H,\le)
		\label{HA_algebraic_struct}
	\end{align}
	Where:\begin{itemize}
		\item \(X=H(G \cup C)\) is the universe of discourses containing all possible linguistic labels.
		\item \(G=\{g^-,g^+\}=\{\text{\textit{small}},\text{\textit{big}}\}\) is the set of primitive linguistic terms (i.e, generators \cite{HO2008968}).
		\item \(C=\{0,W,1\}\) where \(W=\text{\textit{neutral}}\), \(0=\text{\textit{absolutely small}}\), \(1=\text{\textit{absolutely big}}\) are constants.
		\item \(H=H^+ \cup H^-\) where \(H^+=\{\text{\textit{very}}\}\) is the set of all chosen positive hedges and \(H^-=\{\text{\textit{little}}\}\) is the set of all chosen negative hedges. Hedges are unary operators that either increase or decrease semantic degree of a linguistic term.
		\item \(\le=\{\text{\textit{very small}}, \text{\textit{small}}, \text{\textit{little small}}, \text{\textit{neutral}}, \text{\textit{little big}}, \\\text{\textit{big}}, \text{\textit{very big}}\}\) is the partially semantics-based ordering relation on \(X\), holding all chosen linguistic labels in the order that increases the labels' semantic meanings from left to right \cite{HO2008968}.
	\end{itemize}
	
	The semantically quantifying mapping (SQM) $\nu:X\rightarrow [0,1]$ is defined below:
	\begin{align}
		\nu(W)&=fm(g^-)=\theta\\
		\nu(g^-)&=\theta-\alpha\,fm(g^-)=\beta\,fm(g^-)\\
		\nu(g^+)&=\theta+\alpha\,fm(g^+)\\
		\begin{split}
			\nu(h_j\,x)=\nu(x)&+Sign(h_j\,x)\,[\sum_{i=Sign(j)}^{j}fm(h_i\,x) \\
			&-\omega(h_j\,x)fm(h_j\,x)]
		\end{split}\label{SQM_equ}
	\end{align}
	where \(fm(\cdot)\) is a function that measures fuzziness, \(Sign(\cdot)\) is a redefined \(Sign\) function for HA application, \(j\in \{j:-q\le j \le p \text{ and }j\not =0\}\), \(\alpha=\sum_{i=-q}^{-1}\mu(h_i)\) is the sum of fuzziness measures of all negative hedges, \(\beta=1-\alpha\) is the sum of fuzziness measures of all positive hedges and \(\omega(h_j\,x)=0.5\,[1+Sign(h_j\,x)Sign(h_p\,h_j\,x)(\beta-\alpha)]\).
	
	\underline{\textbf{Example:}} The SQM values of the linguistic variable \textbf{SIZE} whose the HA structure is described in equation \ref{HA_algebraic_struct} can be calculated symbolically as shown in table \ref{SQM_table}.
	\begin{table}[h]
		\centering
		\normalsize
		\caption{SQM values of SIZE}
		\label{SQM_table}
		\begin{tabularx}{0.8\textwidth}{*{2}{Y}}
			\toprule
			\textbf{Labels} & \textbf{SQM values}                   \\ \midrule
			Very small      & \(\theta(1-\alpha)^2\)                \\
			Small           & \(\theta(1-\alpha)\)                  \\
			Little small    & \(\theta(1-\alpha+\alpha^2)\)         \\
			Neutral         & \(\theta\)                            \\
			Little big      & \(\theta+\alpha(1-\theta)(1-\alpha)\) \\
			Big             & \(\theta+\alpha(1-\theta)\)           \\
			Very big        & \(\theta+\alpha(1-\theta)(2-\alpha)\) \\ \bottomrule
		\end{tabularx}
	\end{table}
	
	\subsection{Hedge algebra controller structure}\label{HAC_stages}
	There are many ways in which a HAC can be designed for a particular system. The usual structure is presented in \cite{HO2008968}, which has been successfully applied to solve various real-world problems such as in \cite{HAC_earthquake_2013}. However, in this section, we will generalize a more straightforward, efficient structure that has been developed in \cite{intermediate_ctrl_2022}. Although this structure is first applied to solve the cart-pole problem, it can be generalized to use in other systems. The overall structure are demonstrated in figure \ref{HAC_struct}.
	\drawHACgeneralStructure
	
	Particularly, the stages included in the structure are:
	\begin{enumerate}[\text{Stage} 1.]
		\item \textit{Semantization} is a transformation from the crisp (real) domain \([a,b]\) of a variable to its semantic domain \([a_s,b_s]\subseteq [0,1]\). Below is a linear semantization that transforms variable \(x\in [a,b]\) to \(x_s\in [a_s,b_s]\).
		\begin{align}
			x_s=a_s+(x-a)\frac{b_s-a_s}{b-a}\label{linear_sem}
		\end{align}
		It is necessary to note that when \([a_s,b_s]=[0,1]\) equation \ref{linear_sem} become a simple normalization \(x_s=\frac{x-a}{b-a}\). It is also possible to use a nonlinear mapping for semantization \(x_s=f(x)\), as long as \(f(x)\) is bijective, continuous and monotonically increasing.
		
		\item \textit{Inference} is an interpolation process that infers intermediate control actions for each system's state in the semantic domains. For instance, consider variable \(x_s\) having these linguistic labels \{very negative, negative, little negative, neutral, little positive, positive, very positive\}, we can calculate SQM values of these labels using equation \ref{SQM_equ} to get a list of SQM values of \(x_s\). The same process can be done for \(u_s\) to get a list of SQM values of \(u_s\). We use these two SQM value lists of \(x_s\) and \(u_s\) to construct an inference mechanism as an interpolation problem between the two lists to deduce \(u_s\) from \(x_s\) for each system's state.
		
		\item \textit{De-semantization} is the inverse mapping of \textit{semantization} that maps each intermediate control action in the semantic domain to its crisp (real) domain.
		
		\item \textit{Combination of weighted intermediate outputs} is where we assign weights to each intermediate control action and sum (or combine) them linearly (or non-linearly) to get the final control action(s). The design of this stage depends on the number of system's inputs and outputs, the control criteria and the desire behaviors of the controller. The weights can be static or adaptive to the current states of the system as shown in \cite{intermediate_ctrl_2022}.
	\end{enumerate}
	
	\section{Hedge algebra controllers with recursive semantic values}\label{RSHAC}
	In order to further refine theoretical bases of hedge algebra and overcome all the drawbacks of hedge algebra based controllers mentioned in section \ref{intro}, this paper proposes additional concepts that can be incorporated in the same aforementioned structure of a HAC.
	
	\subsection{Infinite general semantization/de-semantization}
	In section \ref{HAC_stages}, it can be seen that the semantization process requires knowledge about the variable's crisp domain \([a,b]\). However, this insight sometimes cannot be determined in prior to the design of controllers. For example, consider under-actuated systems such as a cart-pole, even though the limits of the cart's position and cart's velocity can be pre-determined, the range of the pendulum's angular velocity is unknown because the pendulum is not actuated. In such a case, a semantization mapping from \((-\infty,\infty)\) to \((0,1)\) is needed. Therefore, we propose the infinite general semantization $IGS:(-\infty,\infty)\rightarrow (0,1)$ as follows:
	\begin{figure}[!t]
		\centering
		\includegraphics[width=0.9\textwidth,keepaspectratio=true]{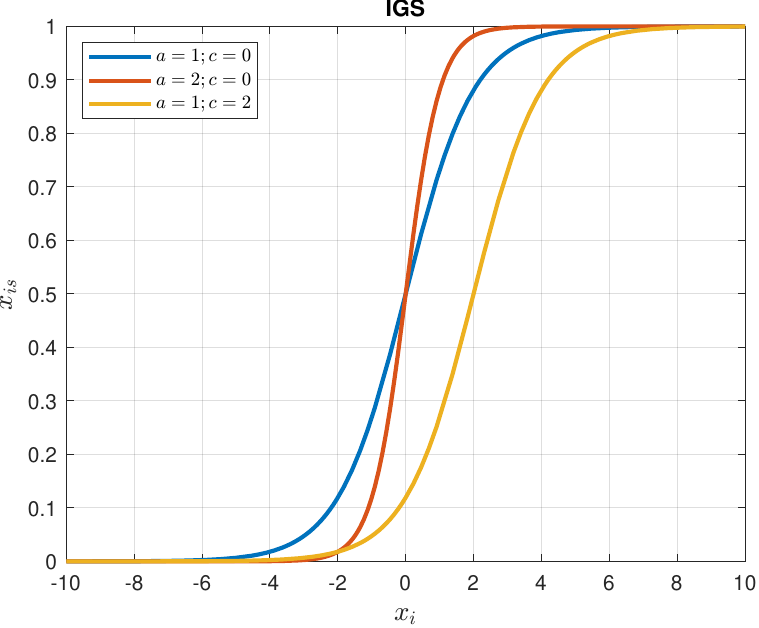}
		\caption{Infinite general semantization (IGS) plotted with different parameters.}\label{IGS_fig}
	\end{figure}
	\begin{align}
		x_{s}=IGS(x)=\frac{1}{1+e^{-a\,(x-c)}}\label{IGS}
	\end{align}
	Where \(a,\,c\in \mathbb{R}\) (such that \(a\ge 0\)) are tunable parameters.
	
	For de-semantization, its inverse mapping called infinite general de-semantization $IGDS:(0,1)\rightarrow (-\infty,\infty)$ can be easily derived as follows:
	\begin{align}
		x=IGDS(x_{s})=\frac{1}{a}\,\ln(\frac{1-x_{s}}{x_{s}})+c\label{IGDS}
	\end{align}
	
	IGS has the following properties:
	\begin{enumerate}
		\item \( \lim_{x \to c} IGS(x) = 0.5\)
		\item \( \lim_{x \to +\infty} IGS(x) = 1\)
		\item \( \lim_{x \to -\infty} IGS(x) = 0\)
		\item \( IGS'(x) = a\,IGS(x)\,(1-IGS(x))\)
	\end{enumerate}
	
	\underline{\textbf{Remark:}} It can be seen from figure \ref{IGS_fig} that this semantization is an one-to-one mapping from \((-\infty,\infty)\) to \((0,1)\), and it is independent of the crisp domain of the variable \(x\). Therefore, it can be applied in both cases of unknown crisp domain, and known crisp domain. Additionally, there are also two tunable parameters which are \(a\) for adjusting the slope and \(c\) for the offsetting horizontally from \(0\). Therefore, for a system's state that is symmetric about \(0\), \(c\) should be equal to \(0\), otherwise different from \(0\).
	
	\subsection{Semantically quantifying simplified mapping}
	In this section, we introduce a new method of quantifying the semantic meaning of a linguistic label (term) called "semantically quantifying simplified mapping" (SQSM). Consider the set \(\le\) from equation \ref{HA_algebraic_struct}, let us define the following concepts.
	
	\begin{itemize}
		\item \textit{Semantic index extraction (SIE)} \(SIE:X\rightarrow \{1,2,...,n_{\le}\}\) returns the index of the linguistic label \(x\in \le\) starting from \(1\).
		\begin{align}
			IDX_x=SIE(x)
		\end{align}
		Where \(n_{\le}=length(\le)\) and is an odd number because there is always a neutral linguistic term.
		\item\textit{ Median semantic index (MSI)} is the index of the neutral element \(W\) in the set \(\le\) starting from \(1\).
		\begin{align}
			MSI=\frac{n_{\le}-1}{2}+1
		\end{align}
		\item \textit{Semantically quantifying simplified mapping (SQSM)} \(\nu:X\rightarrow [0,2\,\theta]\).
		\begin{align}
			\nu(x)=\begin{cases}
				\theta(1-\alpha^{SIE(x)})&; SIE(x)<MSI\\
				\theta(1+\alpha^{n_{\le}+1-SIE(x)})&; SIE(x)>MSI\\
				\theta&; otherwise\\
			\end{cases}
		\end{align}
		Where \(\theta=fm(g^-)\) and \(\alpha\) are tunable parameters. Both of these should be chosen in \((0,1)\).
	\end{itemize}
	
	\begin{algorithm}[h]
		\caption{Generate SQSM values recursively}\label{SQSM_algorithm}
		\KwData{\(n_\le \geq 1\); \(\theta \in [0,1]\); \(\alpha \in [0,1]\)}
		\KwResult{List of \(n_\le\) SQSM values}		
		\(MSI \gets \frac{n_{\le}-1}{2}+1\);
		
		\(SQSMs \gets zeros(n_\le)\);
		
		\(SQSMs[MSI] \gets \theta\);
		
		\(i \gets 1\);
		
		\While{\(i < MSI\)}{
			\(SQSMs[i] \gets \theta(1-\alpha^{i})\);
			
			\(SQSMs[n_\le + 1 - i] \gets \theta(1+\alpha^{i})\);
			
			\(i \gets i+1\);
			
		}
	\end{algorithm}
	
	These novel concepts SIE, MSI, and SQSM are specifically designed to be recursively programmable. Essentially, the semantic degree of a linguistic term can be automatically and programmatically calculated without human's knowledge about the influence of hedges elements on the primitives linguistic terms. Shown in algorithm \ref{SQSM_algorithm}, the proposed algorithm is presented in form of pseudocode with the complexity order of \(O(\frac{n}{2})\).
	
	\underline{\textbf{Example:}} Applying algorithm \ref{SQSM_algorithm} to the same example from section \ref{HA_theory}, we get the SQSM values of the linguistic label \textbf{SIZE} calculated symbolically as shown in table \ref{SQSM_table}.
	\begin{table}[h]
		\centering
		\normalsize
		\caption{SQSM values of SIZE (4\(^*\) is the MSI)}
		\label{SQSM_table}
		\begin{tabularx}{0.8\textwidth}{*{4}{Y}}
			\toprule
			\textbf{Labels} & \textbf{SIE(x)} & \multicolumn{2}{c}{\textbf{SQSM values}} \\ \midrule
			Very small      & 1               &  \multicolumn{2}{c}{\(\theta(1-\alpha)\)}  \\
			Small           & 2               & \multicolumn{2}{c}{\(\theta(1-\alpha^2)\)} \\
			Little small    & 3               & \multicolumn{2}{c}{\(\theta(1-\alpha^3)\)} \\
			Neutral         & 4\(^*\)           &       \multicolumn{2}{c}{\(\theta\)}       \\
			Little big      & 5               & \multicolumn{2}{c}{\(\theta(1+\alpha^3)\)} \\
			Big             & 6               & \multicolumn{2}{c}{\(\theta(1+\alpha^2)\)} \\
			Very big        & 7               &  \multicolumn{2}{c}{\(\theta(1+\alpha)\)}  \\ \bottomrule
		\end{tabularx}
	\end{table}
	
	\underline{\textbf{Remark:}} The main advantage of SQSM over the old SQM from section \ref{HA_theory} is that SQSM has a recursive, programmable definition that allows the automatic generation of SQSM values using algorithm \ref{SQSM_algorithm}. This is impossible when using SQM because, from table \ref{SQM_table}, it can be observed that each SQM formula is unique for its corresponding label. The reason for this is because SQM uses the \(Sign\) function defined in \cite{HO2008968}. As a result, SQM needs human assessment of how the hedge elements (such as \textit{very} and \textit{little} in the example) semantically influence the primitive terms (\textit{big} and \textit{small} in the example) to formulate the SQM values for each label. In the case of SQSM, we use an auto-generated semantic index list as shown in the second column of table \ref{SQSM_table} to maintain the semantic order. Therefore, SQSM does not require manual assessment to formulate. Thus, it is quicker and easier to design, especially when the number of linguistic labels is large. For example, suppose a linguistic variable has 99 labels. In that case, it may take significant time and effort to manually derive SQM formulas for all 99 labels, while SQSM requires no more than a while-loop in any programming language. Therefore, SQSM allows controllers to be scaled easily to high-dimensional systems. Another advantage of SQSM over SQM is that SQSM allows dynamically changing the number of linguistic labels in the middle of the controller's operation because generating new SQSM values can be done programmatically. In contrast, in the case of SQM, the same process is impossible to conduct without manually re-formulating the whole array of SQM values. Therefore, the number of SQSM linguistic labels can also be adjusted during optimization. Inherently, SQSM allows controllers to be more customizable and much less time-consuming to tune.
	
	\subsection{The structure of RS-HAC}\label{RSHAC_stages}
	This section illustrates how to incorporate these new concepts into the HAC's structure in section \ref{HAC_stages}. The new version of HAC is called "Hedge algebra controllers with recursive semantic values" (RS-HAC).
	
	Following the same HAC structure shown in figure \ref{HAC_struct}, the stages of a RS-HAC's structure are:
	\begin{enumerate}[\text{Stage} 1.]
		\item \textit{Semantization}
		
		Although the infinite general semantization (defined by equation \ref{IGS}) can also be used to semantize bounded states (i.e, crisp domain is pre-determined), it is more efficient to use the linear semantization (defined by equation \ref{linear_sem}) to semantize the state into its chosen semantic domain \([a_s,b_s]\subseteq [0,1]\) because then we have fewer parameters to tune. As a result, the parameter tuning process is more straightforward. Otherwise, the IGS should be used to map the state into the semantic domain of \((0,1)\), and parameters \(a\) and \(c\) are used for tuning. We highly recommend that if the unbounded state is symmetric about \(0\) in its crisp domain, \(c\) should be \(0\) to maintain the symmetricity in the semantic domain.
		\item \textit{Inference} 
		
		The same inference process in section \ref{HAC_stages} is used. However, the fundamental difference is that SQSM is employed more than SQM. Consequently, this leads to one underlying change in the design of this stage. In particular, different interpolation lines are constructed for each system's state, increasing the customizability of the design. Thus, this potentially improves the controller's performance and robustness because different states may have different impacts on the overall performance of the controller, and control criteria may not be the same for every system's state. For example, consider balancing a cart-pole system; the pendulum's angle is less important than the cart's position. The cart's position does not contribute to the dynamic behaviour of the whole system. Therefore, in such a system, separate SQSM interpolation lines for each state are needed in the inference process. Another benefit of using SQSM over SQM is that different interpolation lines can have vastly different numbers of linguistic labels thanks to the recursive nature of SQSM as shown in algorithm \ref{SQSM_algorithm}.
		\item \textit{De-semantization} 
		
		This stage is the same as illustrated in section \ref{HAC_stages}. Nonetheless, it is necessary to note that the newly introduced IGDS can be used to map an intermediate control action to an unbounded crisp domain \((-\infty,\infty)\). In contrast, the inverse mapping of the linear semantization should be used in the case of bounded crisp domains.
\item \textit{Combination of weighted intermediate outputs}
		\par
		This stage calculates an adaptive weighted combination of all intermediate control signals to get the final control action(s). This choice is because we want the RS-HAC to be even more customizable. Another reason is that frequently, a real system may have multiple operating regions (or points) among which its dynamics can differ significantly, as well as the control criteria may not be the same in different operating scenarios, so the weight of each intermediate control signal should be adaptable to changes in operating conditions.
\end{enumerate}

\section{\underline{Application example:} Balancing a cart-pole}\label{app_exp}
	A cart-pole system has been a famous benchmark for control algorithms because of its non-linear, under-actuated and unstable properties. Although a dynamical model of the plant is not needed to design RS-HAC, it is briefly presented below for reference.
	\subsection{Cart-pole mathematical model}\label{cartpole_model}
	Oftentimes, the cart-pole model such as the one in \cite{intermediate_ctrl_2022} is widely used. It has one input being a force \(u=F\) acting on the cart and four outputs being the system's states \(X=\begin{bmatrix}x&\dot{x}&q&\dot{q}\end{bmatrix}^T\) where \(x\) is the cart's position, \(\dot{x}\) is the cart's velocity, \(q\) is the angle deviation of the pendulum from the upright position and \(\dot{q}\) is the angular velocity of the pendulum. Despite the whole system being under-actuated, the cart on which the control force exerts is fully actuated. Therefore, we can directly cancel the cart's dynamics by choosing the input to be the cart's acceleration, meaning \(u=\ddot{x}\). This is called "partial feedback linearization" as described in \cite{underactuated}. In form of state space representation, according to first principle modeling, the dynamics of the cart-pole illustrated in figure \ref{cartpole_fig} is:
	\begin{figure}[!h]
		\centering
		\includegraphics[width=0.9\textwidth,keepaspectratio=true]{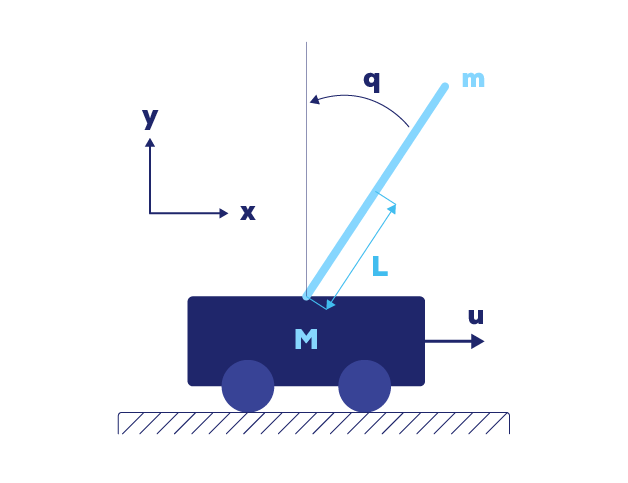}
		\caption{A cart-pole system}
		\label{cartpole_fig}
	\end{figure}
	\begin{align}
		\dot{X}=\begin{bmatrix}
			\dot{x}\\
			\ddot{x}\\
			\dot{q}\\
			\ddot{q}
		\end{bmatrix}
		=\begin{bmatrix}
			\dot{x}\\
			u\\
			\dot{q}\\
			\frac{m\,g\,L\,\sin(q)-m\,L\,\cos(q)\,u-k\,\dot{q}}{I+m\,L^2}
		\end{bmatrix}\label{cartpole_full_dynamic}
	\end{align}
	where \(m\) is the pendulum mass, \(L\) is the length from the revolute joint to the pendulum center of gravity, \(I\) is the pendulum moment of inertia taken at its center of gravity, \(g\) is the gravitational acceleration, and \(k\) is the damping coefficient of the revolute joint.
	
	Since the task of balancing the inverted pendulum on a cart is a linear control problem within the vicinity of the system unstable fixed point \(\bar{X}=\begin{bmatrix}*&0&0&0\end{bmatrix}^T\) (\(*\) represents "don't care" condition), equation \ref{cartpole_full_dynamic} can be linearized using Taylor series as follows.
	\begin{align}
		\dot{X}=\begin{bmatrix}
			0&1&0&0\\
			0&0&0&0\\
			0&0&0&1\\
			0&0&\frac{m\,g\,L}{I+m\,L^2}&\frac{-k}{I+m\,L^2}
		\end{bmatrix}
		\,X+\begin{bmatrix}
			0\\
			1\\
			0\\
			\frac{-m\,L}{I+m\,L^2}
		\end{bmatrix}\,u\label{cartpole_linear_dynamic}
	\end{align}
	
	Equation \ref{cartpole_linear_dynamic} can be discretized with the sampling time \(T_s\) to get the linear discrete dynamics about the system unstable fixed point using Euler method as follows.
	\begin{align}
		\begin{split}
			X_{k+1}=\begin{bmatrix}
				1&T_s&0&0\\
				0&1&0&0\\
				0&0&1&T_s\\
				0&0&\frac{m\,g\,L}{I+m\,L^2}\,T_s&1+\frac{-k}{I+m\,L^2}\,T_s
			\end{bmatrix}
			\,X_k\\+\begin{bmatrix}
				0\\
				T_s\\
				0\\
				\frac{-m\,L}{I+m\,L^2}\,T_s
			\end{bmatrix}\,u_k
		\end{split}
		\label{cartpole_discrete_dynamic}
	\end{align}
	The specifications (measured from a real system) are: 
	\begin{itemize}
		\item \(m=0.116527\,kg\)
		\item \(L=0.15\,m\)
		\item \(g=9.80665\,m/s^2\)
		\item \(I=8.7395\times 10^{-4}\,kg.m^2\)
		\item \(k=0.000161\,N.s/rad\)
		\item \(T_s=0.001\,s\)
	\end{itemize}
	
	\subsection{RS-HAC design}\label{RSHAC_cartpole_design}
\begin{figure*}[!ht]
		\centering
		\includegraphics[width=\textwidth,keepaspectratio=true]{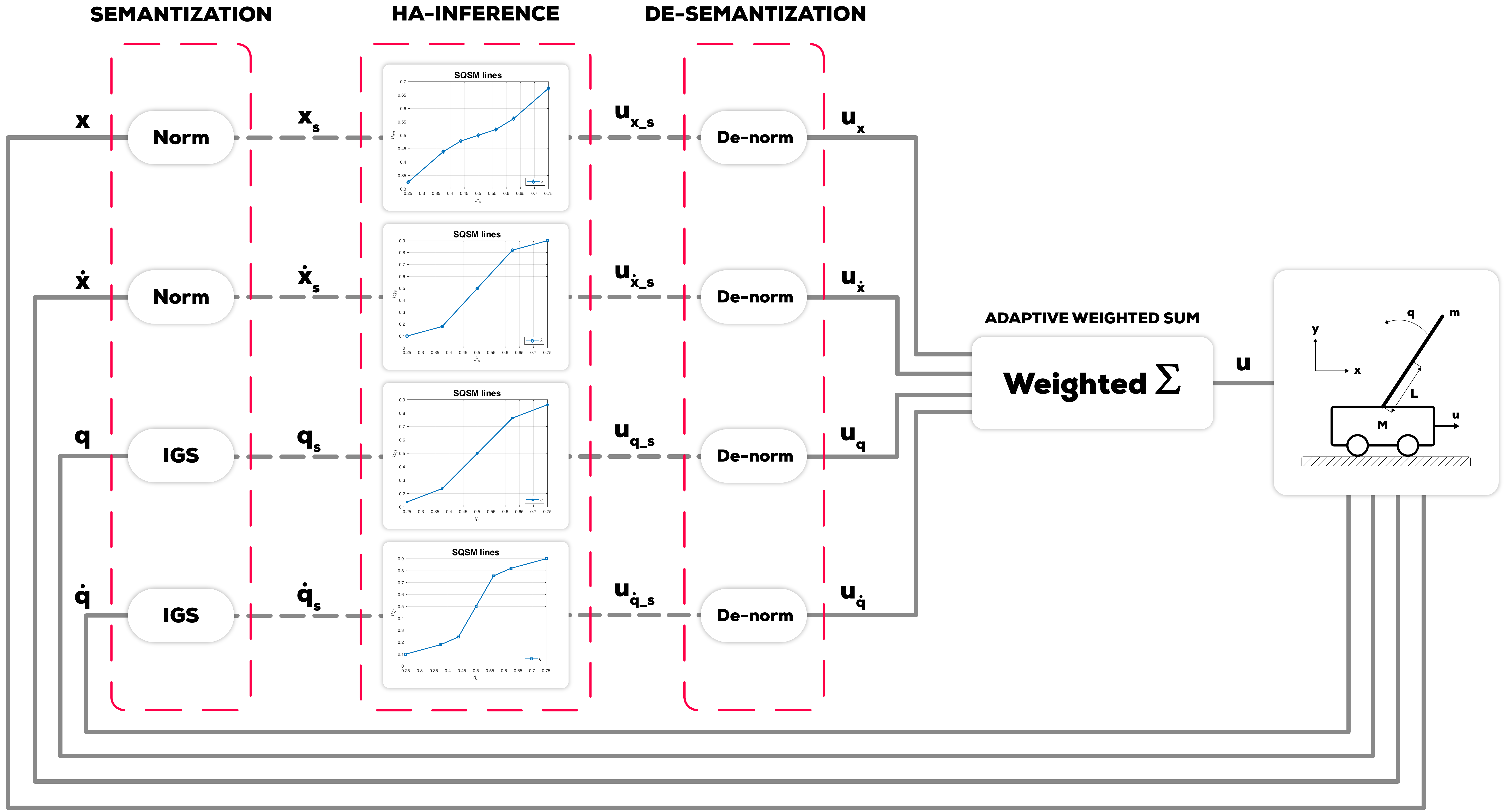}
		\caption{RS-HAC structure for controlling a cart-pole. Dashed lines represent operations in semantic domains, while solid lines illustrate operations in real domains}
		\label{CartPole_RSHAC_struct}
	\end{figure*}

	Illustrated in figure \ref{CartPole_RSHAC_struct}, the RS-HAC stages for this cart-pole plant can be designed as follows.
	
	\begin{enumerate}[\text{Stage} 1.]
		\item \textit{Semantization}
		
		The crisp domains of the cart's position and velocity are pre-determined to be \([-0.43m,0.43m]\) and \([-2m/s,2m/s]\) respectively, while the crisp domain of the pendulum's angle and angular velocity are unbounded. Therefore, we design the semantization stage as shown below.
		\begin{align}
			x_s&=Normalization(x)=\frac{x-x_{min}}{x_{max}-x_{min}}\\
			\dot{x}_s&=Normalization(\dot{x})=\frac{\dot{x}-\dot{x}_{min}}{\dot{x}_{max}-\dot{x}_{min}}\\
			q_s&=IGS(q)=\frac{1}{1+e^{-a_q\,q}}\\
			\dot{q}_s&=IGS(\dot{q})=\frac{1}{1+e^{-a_{\dot{q}}\,\dot{q}}}
		\end{align}
		where \(a_q=8\) and \(a_{\dot{q}}=0.45\) are tuned experimentally, whereas \(c_q=c_{\dot{q}}=0\) to maintain symmetricity.
		\item \textit{Inference} 
		
		Using algorithm \ref{SQSM_algorithm}, we construct 4 different SQSM lists of values for each system's state (\( x, \dot{x}, q, \dot{q}\)) and 4 others for each intermediate control action (\(u_x, u_{\dot{x}}, u_q, u_{\dot{q}}\)) corresponding to the four states. Then, the inference process of the RS-HAC is designed according to the following observations: 
		\begin{enumerate}
			\item If the pendulum angle \(q\) is positive, its intermediate control input is positive, and vice versa. The same rule is applied to \(\dot{q}\) and its intermediate control signal.
			\item If the cart position \(x\) is positive and the pendulum angle is around zero (neutral in semantic domain), the control input is positive to make the pendulum angle \(q\) negative, and then the first rule will incur a negative control input to make the pendulum slowly come back to the inverted equilibrium at the same time as the cart position goes to zero.
		\end{enumerate}
		\begin{figure}[!h]
			\centering
			\includegraphics[width=0.9\textwidth,keepaspectratio=true]{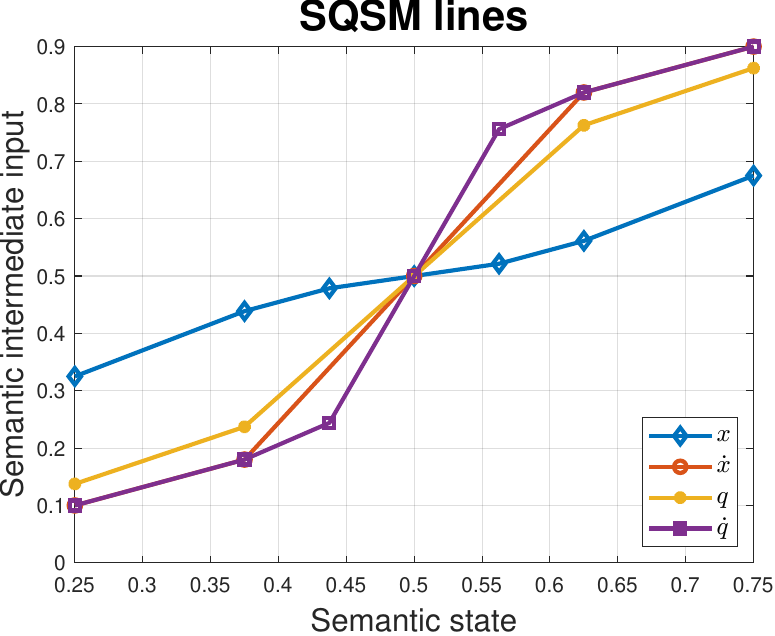}
			\caption{SQSM interpolation lines for each states}
			\label{SQSM_lines}
		\end{figure}
		These rules can be represented in 4 SQSM interpolation lines shown in figure \ref{SQSM_lines} corresponding to each separate inference in the semantic domain (\(x\rightarrow u_x\), \(\dot{x}\rightarrow u_{\dot{x}}\), \(q\rightarrow u_q\), \(\dot{q}\rightarrow u_{\dot{q}}\)). The parameters for these lines (tuned heuristically) is shown in table \ref{SQSM_params_table}. It is noted that the number of linguistic labels \(n_{\le}\) is an odd number for there is a neutral label.
		\begin{table}[!h]
			\centering
			\caption{SQSM parameters for each state and its intermediate controller}
			\label{SQSM_params_table}
			\begin{tabularx}{0.9\textwidth}{cc *{8}{Y}}
				\cmidrule(l){3-10}
				&& \(x\)   & \(u_x\)  & \(\dot{x}\) & \(u_{\dot{x}}\) & \(q\)   & \(u_q\)   & \(\dot{q}\) & \(u_{\dot{q}}\)\\
				\cmidrule(r){1-2}\cmidrule(lr){3-4}\cmidrule(lr){5-6}\cmidrule(lr){7-8}\cmidrule(l){9-10}
				\multicolumn{2}{c}{\(n_{\le}\)} & \(7\)   & \(7\)    & \(5\)       & \(5\)           & \(5\)   & \(5\)     & \(7\)       & \(7\)\\
				\multicolumn{2}{c}{\(\alpha\)} & \(0.5\) & \(0.35\) & \(0.5\)     & \(0.8\)         & \(0.5\) & \(0.725\) & \(0.5\)     & \(0.8\)\\
				\multicolumn{2}{c}{\(\theta\)} & \(0.5\) & \(0.5\)  & \(0.5\)     & \(0.5\)         & \(0.5\) & \(0.5\)   & \(0.5\)     & \(0.5\)\\
				\bottomrule
			\end{tabularx}
		\end{table}
		\item \textit{De-semantization}
				
		Since the input \(u\) in this case is bounded, the inverse mapping of equation \ref{linear_sem} is used.
		\begin{align}
			u_i=u_{is}\,(u_{max}-u_{min})+u_{min}
		\end{align}
		Where \(i\in \{x,\dot{x},q,\dot{q}\}\) and \(u_{max}=29.42\,m/s^2\), \(u_{min}=-29.42\,m/s^2\).
		\item \textit{Combination of weighted intermediate control signals}
		
		The intermediate actions are weighted, using the same technique shown in \cite{intermediate_ctrl_2022}. These weights are calculated adaptively, depending on the current absolute value of the pendulum angle (\(|q|\)).
		\begin{align}
			&\text{If \(|q| \le l_1\), }w_{q}=w_{\dot{q}}=w_x=w_{\dot{x}}=0.25\\
			&\text{If \(l_1 < |q| < l_2\), }\begin{cases}
				w_{q}&=0.25 + (q-l_1)\frac{1-0.25}{l_2-l_1}\\
				w_{\dot{q}}&=\frac{1-w_{q}}{2}\\
				w_x&=w_{\dot{x}}=\frac{1-w_{q}-w_{\dot{q}}}{2}
			\end{cases}\\
			&\text{If \(|q| \ge l_2\), }w_{q}=1, w_{\dot{q}}=w_x=w_{\dot{x}}=0
		\end{align}
		Where \(0<l_1<l_2<\frac{\pi}{2}\). These parameters (\(l_1=0.09\,rad\) and \(l_2=0.87\,rad\)) are tuned heuristically. The final input action is the sum of all weighted intermediate control signals.
		\begin{align}
			u=\sum_{j\in \{x,\dot{x},q,\dot{q}\}} w_j\,u_j
		\end{align}
	\end{enumerate}
	\subsection{Experimental procedure}
	The RS-HAC is compared with a LQR (Linear quadratic regulator) and a FC (fuzzy controller) in two simulated experiments regarding balancing an inverted pendulum on a cart. The FC and LQR are established as shown below.
\subsubsection*{Reference controllers design}
	\begin{itemize}
		\item[\underline{FC:}] \textit{Fuzzy controller}
		\begin{figure}[H]
			\centering
			\includegraphics[width=0.9\textwidth, keepaspectratio=true]{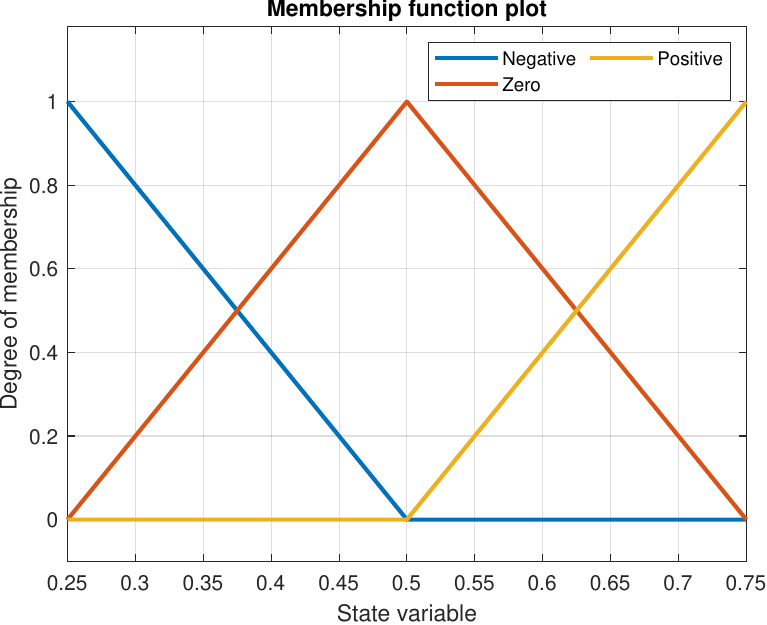}
			\caption{Membership functions for each state variable}\label{MF_FC}
		\end{figure}	
		\begin{figure}[H]
			\centering
			\includegraphics[width=0.9\textwidth, keepaspectratio=true]{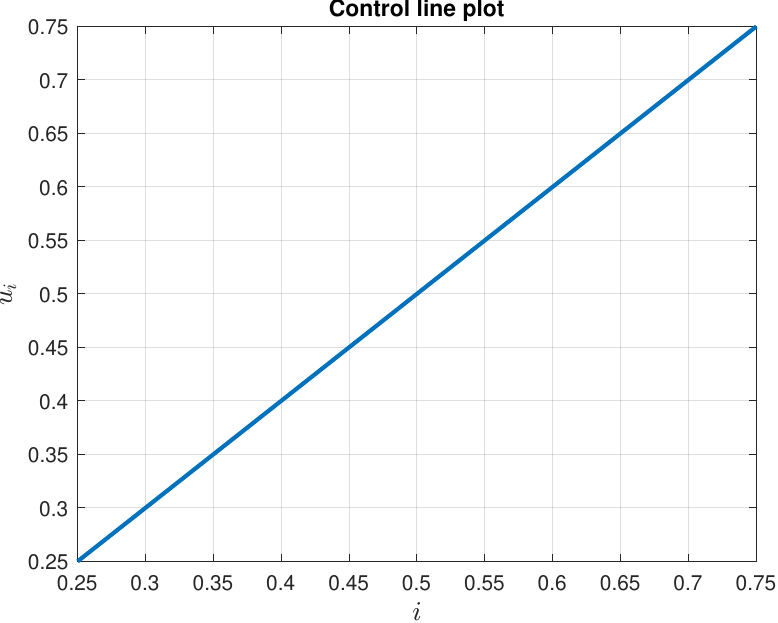}
			\caption{Fuzzy control line where \(i\in \{x,\dot{x},q,\dot{q}\}\)}\label{fuzzy_inference}
		\end{figure}
		For fairness, the FC is designed to have the exact same structure as the RS-HAC which is shown in figure \ref{CartPole_RSHAC_struct}, except for the second stage (HA inference) being replaced with a type-1 Takagi-Sugeno fuzzy inference system \cite{6313399}, meaning that stage 1, stage 3 and stage 4 in section \ref{RSHAC_cartpole_design} are reused for FC. Stage 2 is a fuzzy inference system having 4 inputs (4 state variables in fuzzy domain \([0, 1]\)) and 4 outputs (4 intermediate control signals in fuzzy domain \([0, 1]\)). The single input rule modules (SIRMs) demonstrated in the paper \cite{YI2000153} is used to design this inference stage. Essentially, there are 3 triangular membership functions plotted in figure \ref{MF_FC} (corresponding to 3 linguistic labels "Negative", "Zero" and "Positive") for each state variable. Following SIRMs, the rules are made separately for each state variable according to the same observations in stage 2 of RSHAC in section \ref{RSHAC_cartpole_design}. Finally, the intermediate control signal of each input is given via a weighted average de-fuzzification process, which makes the control surface to be a straight line plotted in figure \ref{fuzzy_inference}.
		\item[\underline{LQR:}] \textit{Linear quadratic regulator}
		
		LQR is a widely popular model-based optimal controller to balance the cart-pole \cite{underactuated}. It is a gain controller whose gain is computed as the solution of an offline optimization. The behavior of this controller depends on the state's penalties \(Q\) and the control input penalty \(R\) in the objective function of the offline optimization. With the system dynamics shown in equation \ref{cartpole_discrete_dynamic} and the choices for \(Q\) and \(R\) as shown in equation \ref{LQR_params}, the LQR gain is calculated in equation \ref{LQR_gain}.
		\begin{align}
			Q&=\begin{bmatrix}
				40 & 0 & 0   & 0 \\
				0  & 1 & 0   & 0 \\
				0  & 0 & 100 & 0 \\
				0  & 0 & 0   & 2
			\end{bmatrix},~R=2\label{LQR_params}\\
			K_{LQR}&=\begin{bmatrix}
				-13.95&-11.69&-56.16&-7.89
			\end{bmatrix}\label{LQR_gain}
		\end{align}
	\end{itemize}
	
	\subsubsection*{Experimental procedure}
	The experimental procedure consists of 2 experiments.
	\begin{enumerate}
		\item \textbf{\textit{Balancing the cart-pole:}}
		
		Starting with the  cart-pole initial condition being \(X_0=[0,0,q_0,0]\), all three controllers (RS-HAC, FC and LQR) are required to stabilize the system at \(X_{ref}=[0,0,0,0]\) in three different scenarios where \(q_0= 10\,deg,20\,deg,30\,deg\).
		\item \textbf{\textit{Stepping the cart-pole to a new location:}}
		
		Starting with the  cart-pole initial condition being \(X_0=[0,0,0,0]\), at \(t=1\,s\) all three controllers (RS-HAC, FC and LQR) are required to bring the system to \(X_{ref}=[0.2\,(m),0,0,0]\), meanwhile maintain the steady state of the pendulum. We only conduct one scenario in this experiment because equation \ref{cartpole_discrete_dynamic} indicates that the cart position is irrelevant to the dynamical behavior of the cart-pole.
	\end{enumerate}
	\subsubsection*{Experiment result}
\begin{figure}[H]
		\centering
		\includegraphics[width=\textwidth, keepaspectratio=true]{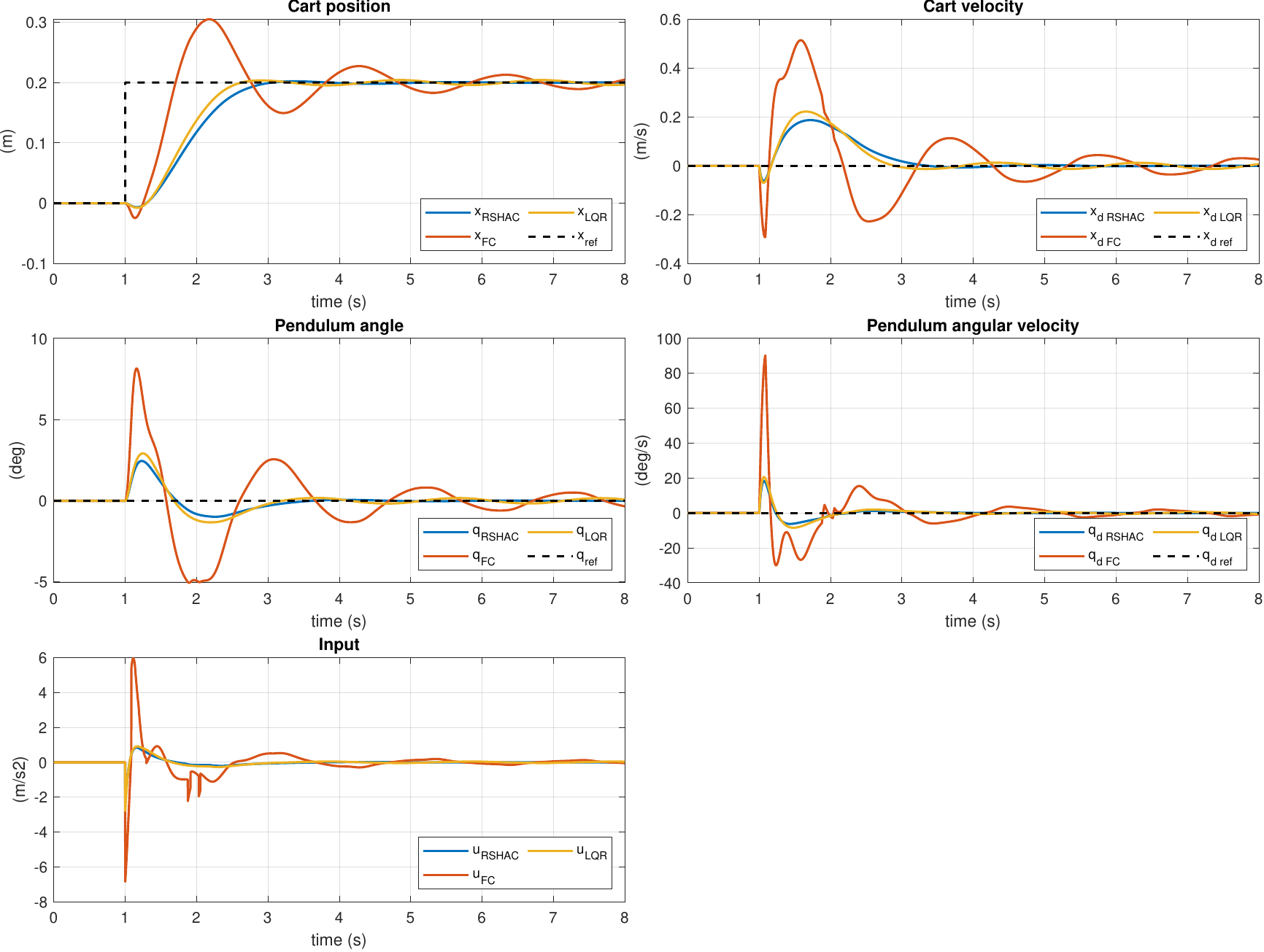}
		\caption{Experiment 2 results}\label{RSHAC_FC_LQR_step}
	\end{figure}

	The results of experiment 1 and 2 are shown in figure \ref{RSHAC_FC_LQR} and \ref{RSHAC_FC_LQR_step} respectively, and summarized in table \ref{exp_sum}. The system is consider stable when the below conditions \ref{stability_cont} are simultaneously satisfied.
	\begin{align}
		\begin{cases}
			|x-x_{ref}|&\le 0.02\,m\\
			|\dot{x}|&\le 0.02\,m/s\\
			|q|&\le 0.5\,deg\\
			|\dot{q}|&\le 0.5\,deg/s\\
		\end{cases}\label{stability_cont}
	\end{align}
	The following performance indices are used. It is noted that the lower these indices are, the better the performance of a controller.
	\begin{enumerate}
		\item \(\Delta\,t\,(s)\) is the transient time, indicating how long it takes a controller to make the system satisfy condition \ref{stability_cont} from \(X_0\).
		\item \(\Delta\,x_{m}\,(m)\) is the maximum position deviation between the reference cart position and the actual cart position during an experiment scenario.
		\item \(\Sigma\,u\,(m/s)\) is the control effort given by equation \ref{ctrl_effort}, where \(k\) is the time step index, \(u_k\,(m/s^2)\) is the control signal at time step \(k\), and \(T_s=0.001\,s\) is the sampling time of the controller.
		\begin{align}
			\Sigma\,u=\sum_{k=0}^{k_{\Delta\,t}}|u_k|\,T_s\label{ctrl_effort}
		\end{align}
		\item For experiment 2 only, \(\%\,x_r\) is used to indicate the overshoot in terms of the cart position.
	\end{enumerate}
	\begin{table}[H]
		\centering
		\scriptsize
		\caption{Experiment summary}
		\label{exp_sum}
		\begin{tabularx}{\textwidth}{cc *{12}{Y}}
			\cmidrule(l){3-14}
			&& \multicolumn{9}{c}{Experiment 1}  
			& \multicolumn{3}{c}{Experiment 2}\\
			\cmidrule(l){3-11} \cmidrule(l){12-14}
			&& \multicolumn{3}{c}{\(q_0=10\,deg\)}& \multicolumn{3}{c}{\(q_0=20\,deg\)}& \multicolumn{3}{c}{\(q_0=30\,deg\)}& \multicolumn{3}{c}{\(x_{ref}=0.2\,m\)}\\
			\cmidrule(l){3-5} \cmidrule(l){6-8} \cmidrule(l){9-11} \cmidrule(l){12-14}
			&& \(\Delta\,t\) & \(\Delta\,x_{m}\) & \(\Sigma\,u\) & \(\Delta\,t\) & \(\Delta\,x_{m}\) & \(\Sigma\,u\) & \(\Delta\,t\) & \(\Delta\,x_{m}\) & \(\Sigma\,u\) & \(\Delta\,t\) & \(\%\,x_r\) & \(\Sigma\,u\)\\
			\cmidrule(r){1-2} \cmidrule(l){3-5} \cmidrule(l){6-8} \cmidrule(l){9-11} \cmidrule(l){12-14}
			\multicolumn{2}{c}{\textbf{RS-HAC}} & \textbf{2.052} &  0.106 &  1.194 & \textbf{2.169} & \textbf{0.187} & 2.835& \textbf{2.598} & \textbf{0.306} & 5.121& \textbf{2.275} & \textbf{1\%} & \textbf{0.495}\\
			\multicolumn{2}{c}{FC}  &  5.238 & \textbf{0.084} &  1.921 &  6.205 & 0.195 &  4.301 & 7.289 & 0.335 & 8.420& 6.136 & 52\% & 2.561\\
			\multicolumn{2}{c}{LQR}  &  2.482 & 0.106 & \textbf{0.996} &  2.704 &  0.219 & \textbf{2.049} & 2.865 & 0.356 & \textbf{3.324} & 2.347 & 2\% & 0.597\\
			\bottomrule
		\end{tabularx}
	\end{table}

	\begin{figure}[H]
		\centering
		\begin{subfigure}{\textwidth}
			\centering
			\includegraphics[width=\textwidth, keepaspectratio=true]{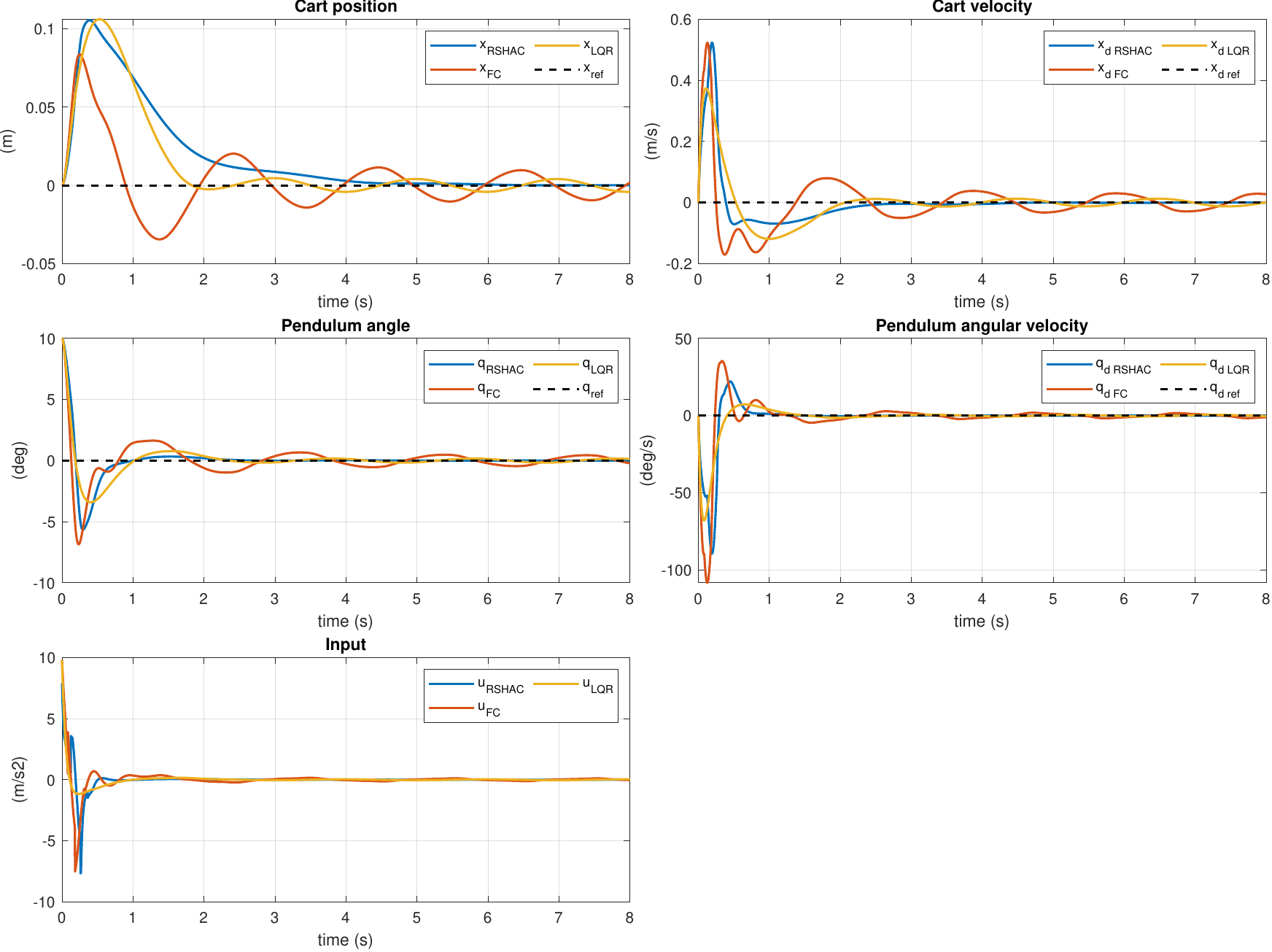}
			\caption{RS-HAC, FC and LQR stabilize a cart-pole (\(X_0=[0,0,10\,(deg),0]^T\))}\label{RSHAC_FC_LQR_10deg}
		\end{subfigure}
	\end{figure}
	\begin{figure}[H]
		\ContinuedFloat
		\begin{subfigure}{\textwidth}
			\centering
			\includegraphics[width=\textwidth, keepaspectratio=true]{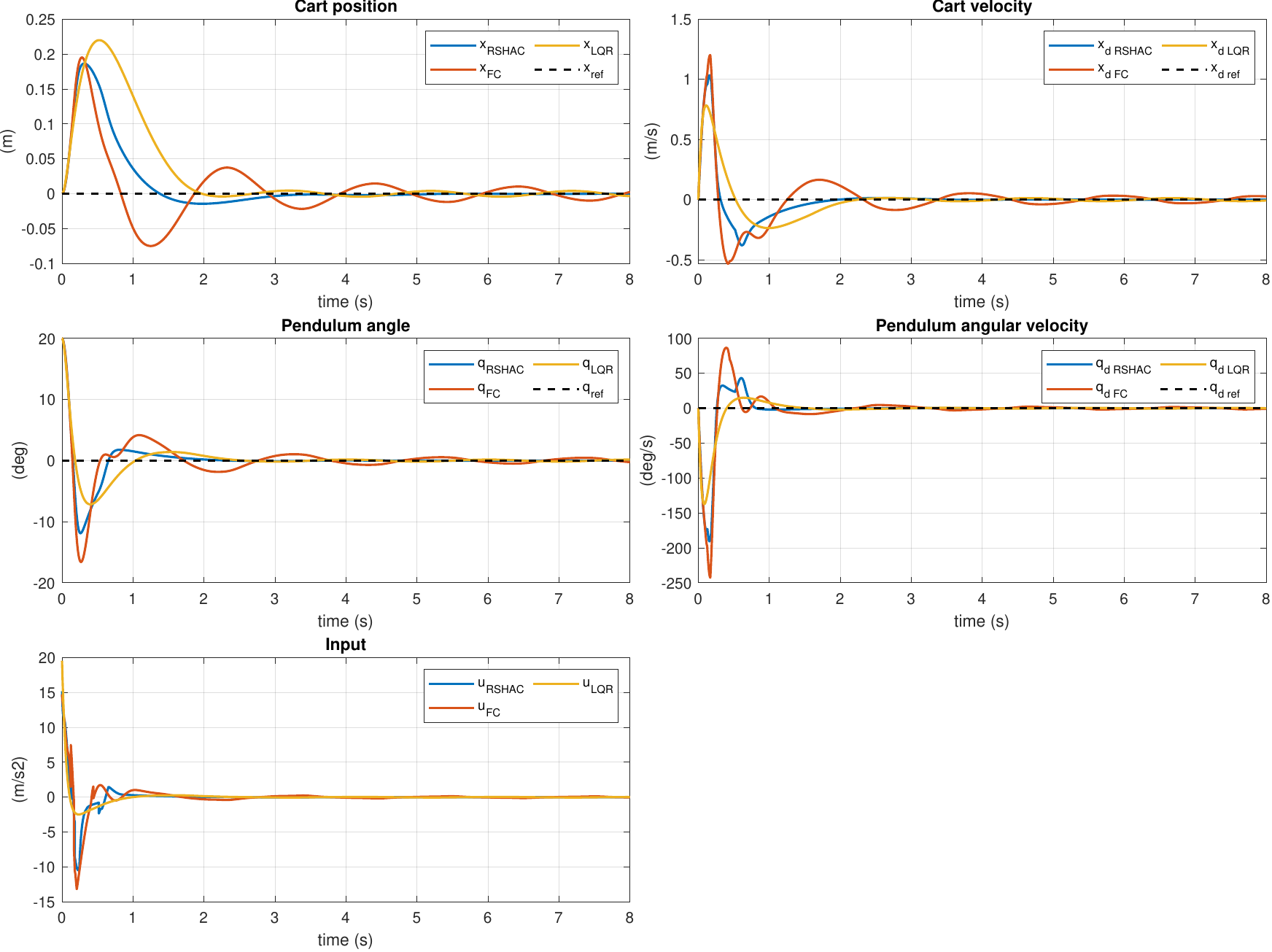}
			\caption{RS-HAC, FC and LQR stabilize a cart-pole (\(X_0=[0,0,20\,(deg),0]^T\))}\label{RSHAC_FC_LQR_20deg}
		\end{subfigure}
	\end{figure}
	\begin{figure}[H]
		\ContinuedFloat
		\begin{subfigure}{\textwidth}
			\centering
			\includegraphics[width=\textwidth, keepaspectratio=true]{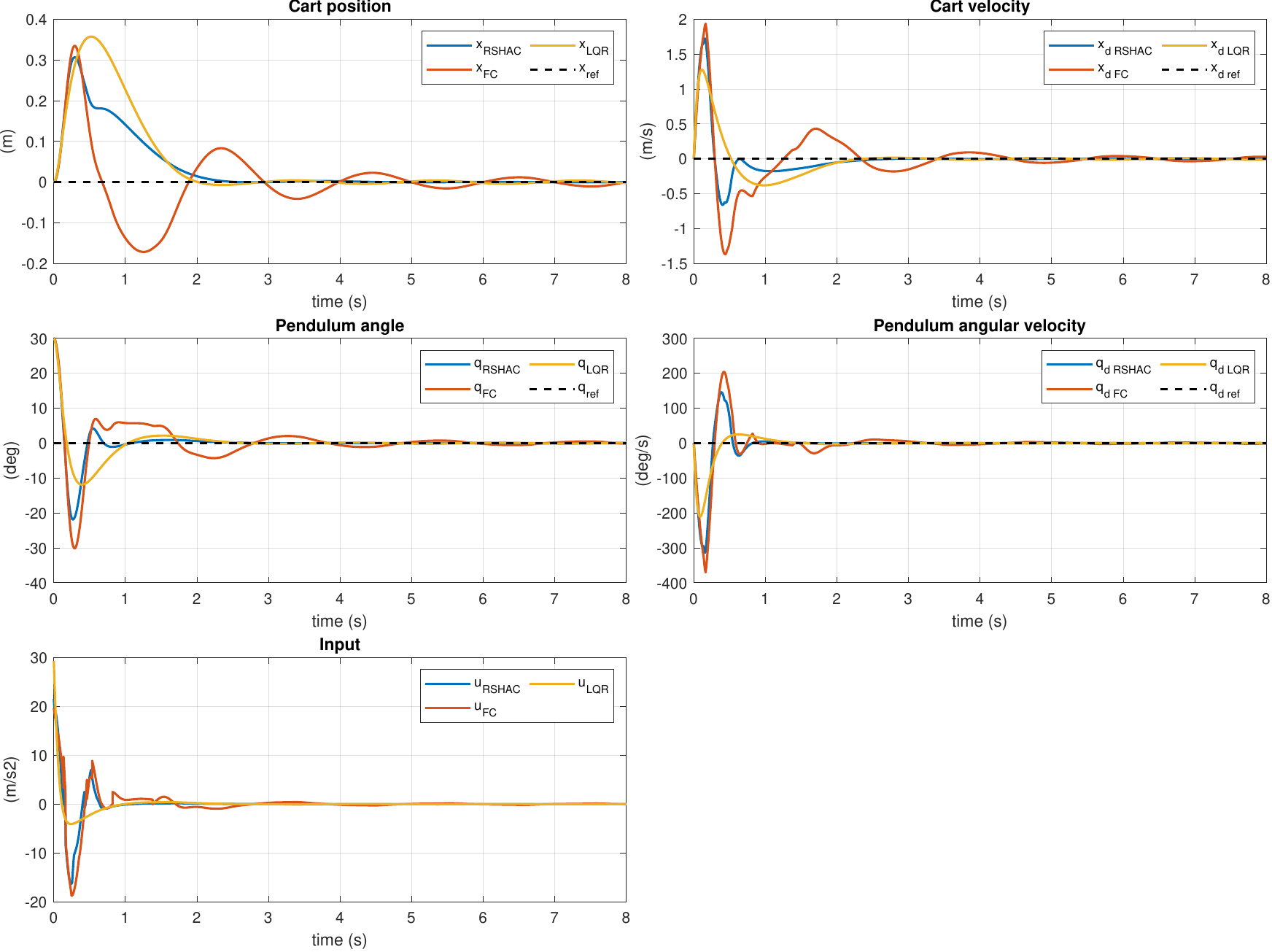}
			\caption{RS-HAC, FC and LQR stabilize a cart-pole (\(X_0=[0,0,30\,(deg),0]^T\))}\label{RSHAC_FC_LQR_30deg}
		\end{subfigure}
		\caption{Experiment 1 results}\label{RSHAC_FC_LQR}
	\end{figure}
	
	\subsubsection*{Result analysis}
	Table \ref{performance_comparison} gives information about how much better (or worse) the RS-HAC performs in comparison with the FC and LQR for each performance index.
	\begin{table}[H]
		\centering
		\scriptsize
		\caption{Performance comparison for each index}
		\label{performance_comparison}
		\begin{tabularx}{\textwidth}{cc *{12}{Y}}
			\cmidrule(l){3-14}
			&& \multicolumn{9}{c}{Experiment 1}  
			& \multicolumn{3}{c}{Experiment 2}\\
			\cmidrule(l){3-11} \cmidrule(l){12-14}
			&& \multicolumn{3}{c}{\(q_0=10\,deg\)}& \multicolumn{3}{c}{\(q_0=20\,deg\)}& \multicolumn{3}{c}{\(q_0=30\,deg\)}& \multicolumn{3}{c}{\(x_{ref}=0.2\,m\)}\\
			\cmidrule(l){3-5} \cmidrule(l){6-8} \cmidrule(l){9-11} \cmidrule(l){12-14}
			&& \(\Delta\,t\) & \(\Delta\,x_{m}\) & \(\Sigma\,u\) & \(\Delta\,t\) & \(\Delta\,x_{m}\) & \(\Sigma\,u\) & \(\Delta\,t\) & \(\Delta\,x_{m}\) & \(\Sigma\,u\) & \(\Delta\,t\) & \(\%\,x_r\) & \(\Sigma\,u\)\\
			\cmidrule(r){1-2} \cmidrule(l){3-5} \cmidrule(l){6-8} \cmidrule(l){9-11} \cmidrule(l){12-14}
			\multicolumn{2}{c}{RS-HAC and FC}  &  \(\uparrow\)155\% & \(\downarrow\)21\% &  \(\uparrow\)61\% &  \(\uparrow\)186\% & \(\uparrow\)5\% &  \(\uparrow\)52\% & \(\uparrow\)181\% & \(\uparrow\)9\% & \(\uparrow\)64\% & \(\uparrow\)170\% & \(\uparrow\)51\% & \(\uparrow\)417\%\\
			\multicolumn{2}{c}{RS-HAC and LQR}  &  \(\uparrow\)21\% & 0\% & \(\downarrow\)17\% &  \(\uparrow\)25\% &  \(\uparrow\)18\% & \(\downarrow\)28\% & \(\uparrow\)10\% & \(\uparrow\)16\% & \(\downarrow\)35\% & \(\uparrow\)3\% & \(\uparrow\)1\% & \(\uparrow\)21\%\\
			\bottomrule
		\end{tabularx}
	\end{table}
	
	Some remarks can be made about these two exercises:
	\begin{enumerate}[\text{Remark} 1.]
		\item In experiment 1, regarding transient time, the RS-HAC is the quickest to stabilize the cart-pole. In particular, during scenario 1 (\(q_0=10\,deg\)), the RS-HAC only takes approximately 2 seconds to stabilize the system. This is 155\% and 21\% faster than the FC and LQR, respectively. The RS-HAC's quick and responsive performance is maintained throughout all three scenarios of experiment 1. However, concerning the control effort, although the RS-HAC are consistently better, being at least 52\% more efficient, than the FC, the LQR controls the cart-pole more effortlessly as can be seen in figure \ref{RSHAC_FC_LQR} where the LQR inputs are much smoother than the other two controllers in all three scenarios. This phenomenon is also observable in table \ref{performance_comparison}, showing that the LQR uses less overall effort to control the system than the RS-HAC. Despite that, the RS-HAC surpasses the LQR in terms of the maximum cart position deviation, being equal to or even better than the LQR by at most 18\%. Oddly, considering \(\Delta\,x_m\), the FC performs relatively better than the RS-HAC in the case where \(q_0=10\,deg\), but it is marginally worse in the other two scenarios where the initial pendulum angle is larger. The reason for this may be that with a slight initial angle, the membership function of the "Zero" linguistic label (figure \ref{MF_FC}) has the most influence on the control actions of the FC since it has the most comprehensive range out of the three membership functions. Thus, when the initial angle is larger, the FC would require more time and space to bring the cart-pole to rest.
\item In experiment 2, looking at figure \ref{RSHAC_FC_LQR_step}, it is straightforward to say that the performances of RS-HAC are almost identical to that of the LQR. This observation is justified by the numbers in table \ref{exp_sum} and \ref{performance_comparison} with only one exception in terms of control effort, which indicates that RS-HAC seems to use slightly less overall effort than LQR does by 21\%. On the other hand, regarding control effort, the RS-HAC is superior to the FC by over 400\%. Additionally, the RS-HAC demonstrates only 1\% overshoot over the reference signal, whereas the FC overshoots the reference by a significant percentage of 52\%. Moreover, the consistency of the RS-HAC in terms of transient time is also shown in experiment 2, where it is remarkably better than that of the FC by 170\%. The RS-HAC is an outstanding controller in terms of transient time, and it is a very suitable controller for stepping the system to another operating location.
\item It can be observed in figure \ref{RSHAC_FC_LQR} that the controllers make the cart positions in all three scenarios rapidly deviate far from 0 at first, then gradually come back to equilibrium. Similarly, in figure \ref{RSHAC_FC_LQR_step}, at \(t=1\,s\), the cart positions of all three controllers slightly move to the opposite direction of the step signal at first, then quickly converge to the new reference location. These phenomena illustrate the successful implementations of the two observations in stage 4 of section \ref{RSHAC_cartpole_design} in which we want to move the cart appropriately to stabilize the pendulum from its initial condition.
		\item Finally, it can be concluded that the RS-HAC is a higher quality controller than the FC, outperforming on almost all performance indices, while its performance is equivalent to that of the LQR.
	\end{enumerate}
	
	\subsection{RS-HAC implementation on real cart-pole system}
	\begin{figure}[H]
		\centering
		\includegraphics[width=0.9\textwidth, keepaspectratio=true]{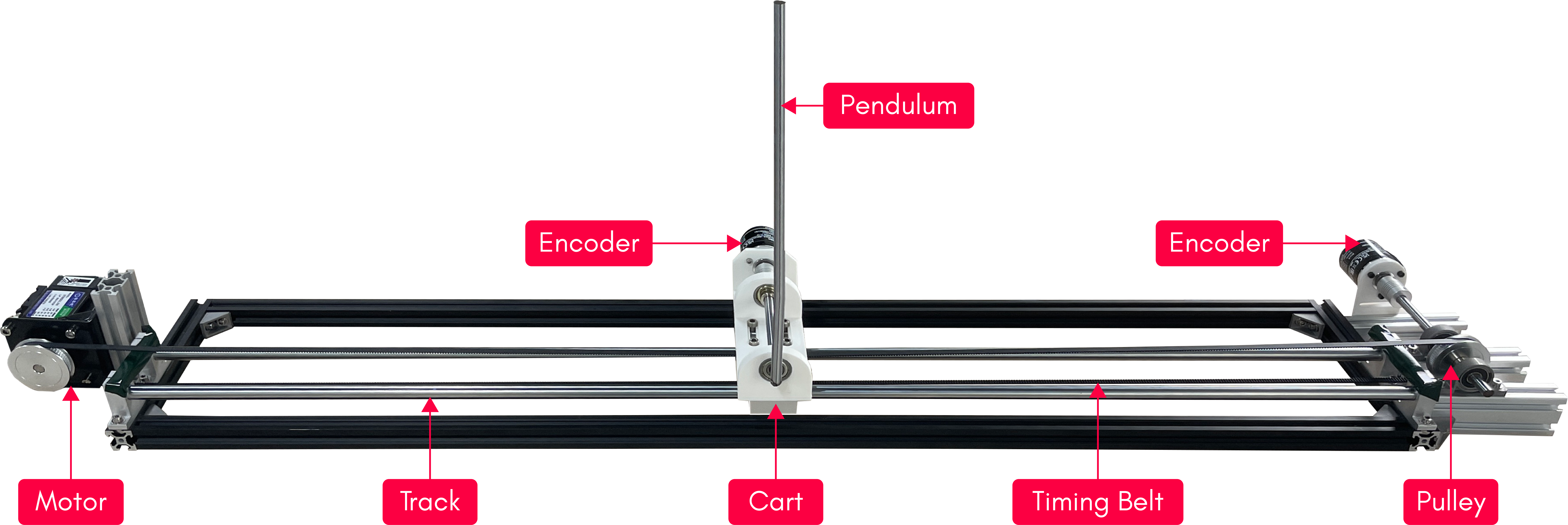}
		\caption{A real inverted pendulum on a cart}\label{real_cartpole}
	\end{figure}
\begin{figure}[H]
		\centering
		\includegraphics[width=\textwidth, keepaspectratio=true]{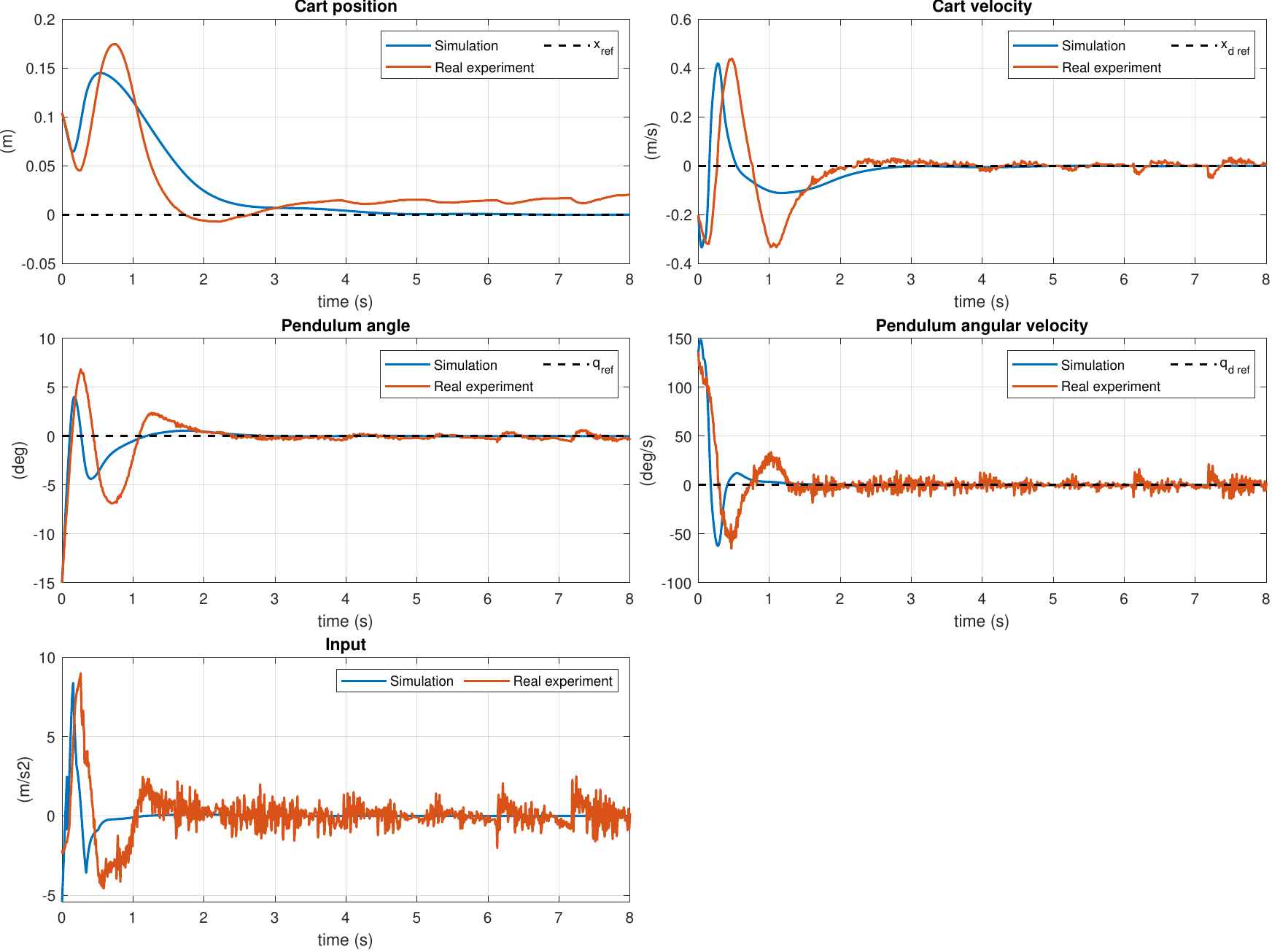}
		\caption{RSHAC balances a real cart-pole in simulation and in real implementation}\label{RSHAC_real}
	\end{figure}

	We implement a RS-HAC on a Teensy 4.1 microcontroller to balance the cart-pole system shown in figure \ref{real_cartpole}. The specification of this system is given at the end of section \ref{cartpole_model}. Because the RS-HAC balance control is only applicable to the linear dynamics of the cart-pole within the vicinity of the unstable inverted fixed point (given by equation \ref{cartpole_linear_dynamic}), we incorporate a swing-up strategy described in \cite{CHATTERJEE2002355} to swing the pendulum from the stable downward fixed point to the region near the inverted fixed point, before activating the RS-HAC to balance the system. This is called a switching controller, and the threshold of the switch is 15 degrees of the pendulum angle. In essence, the RS-HAC only operates whenever \(|q|\le 15\,deg\), otherwise the energy swing-up scheme is used. Therefore, the initial condition of the RS-HAC is whichever system state condition happens to be at the switching point. For example, in the particular case illustrated in figure \ref{RSHAC_real}, the initial condition is \(X_0=[0.10\,(m),-0.20\,(m/s),-14.98\,(deg),135.71\,(deg/s)]^T\), and the reference is \(X_{ref}=[0,0,0,0]^T\).
\par
	Although it can be seen from figure \ref{RSHAC_real} that the simulated and the real RS-HAC share many similarities, such as the general shapes of the corresponding graphs being alike, there are also disparities, including the real RS-HAC lagging behind the simulated by approximately one-fifth of a second, and the input is trembling when the system states close to \(X_{ref}\). The reasons are, first, the real system is not perfectly rigid and has backlashes between assemblies; second, because our actuator is a hybrid servo motor (essentially a stepper motor with the 1.8-degree resolution), it is impossible to have adjustments smaller than 1.8 degrees. Thus, when the states are close to \(X_{ref}\), the motor jumps between \(\pm\)1.8 degrees and 0 degrees, creating the trembling effect.
\section{Conclusion}
This paper introduces new concepts in hedge algebra theory that lay the foundation for a new group of hedge algebra-based controllers. In terms of performance, the proposed controller surpasses the previous class of controllers based on fuzzy inference techniques and can rival the linear quadratic regulator convincingly. These new concepts are invaluable additions to the bases of hedge algebra theory, as they make the theory more diverse and more flexible with the debut of the sigmoid function in the semantization step. More importantly, the newly developed semantically quantifying simplified mapping (SQSM) eliminates all the inherent drawbacks of the SQM, making the design of hedge algebra-based controllers more efficient and straightforward.
\par
Moving forward, we acknowledge that further research is necessary to apply this proposed controller to solve general nonlinear control problems. In addition, incorporating optimization techniques into RS-HAC is pivotal to making this controller a modern control strategy.
	
	\section*{Acknowledgment}
	This research is funded by Hanoi University of Science and Technology (HUST) under project number T2023-TD-017
	\bibliographystyle{elsarticle-num} 
	\bibliography{sources.bib}
\end{document}